\documentclass[twocolumn,prl,floatfix]{revtex4}
\usepackage{graphicx}
\usepackage{times}
\usepackage{color}
\usepackage{amsmath}
\usepackage{epstopdf}

\def\I{{\rm i}}

\def\bs{\boldsymbol}

\begin{document}

\title{Strong Interactions in Crystalline Vacuum}
\author{L.~M. Pismen}
\affiliation{Technion -- Israel Institute of Technology, Haifa 32000, Israel} 

\begin{abstract}
{A multiscale expansion procedure reveals a crystalline vacuum state arising as a result of resonant interactions among weak Planck-scale waves on spinor and cospinor manifolds. Quarks are presented as elastic modulations of the vacuum crystal. The complementing modes are identified with colors, and distinctions between flavors are attributed to asymmetries of the vacuum lattice. Repercussion of the theory include insights into the nature of quantum uncertainty and dark matter and energy.}
\end{abstract}

\maketitle

\textbf{Introduction}. The simple and beautiful structure of the Standard Model (SM) suggests some relation to the structure of vacuum on the Planck scale, which is inaccessible experimentally and can only be probed by its effects on the observable fields on phenomenological scales. This letter explores a mechanism of presenting SM fields as long-scale (``elastic'') modulations of a Planck-scale crystalline structure.

The essential feature of long-scale equations describing envelopes of underlying structures, common in macroscopic nonlinear physics \cite{ch93,lp06}, is their universality, which renders them independent of details of underlying physics and totally determined by the type of the transition and applicable symmetries. Although, contrary to a macroscopic situation, Planck-scale physics is unknown, it is further assumed that the Einstein equations continue to be valid on this scale, adjusted to a topology compatible with the SM structure. The above mentioned resilience of long-scale expansions suggests that the results might be insensitive to deviations from known physics on short scales. The principal tool of the analysis is the multiscale expansion called forth by the huge ``desert'' between the Planck and SM scales. 

It has been long hoped, starting from the original (4+1)D Kaluza--Klein scenario \cite{kk1}, that extra dimensions hold the key to the structure of phenomenological fields localized on the 4D world brane. Compact extra-dimensional manifolds inevitably suffered from transverse instabilities, which were quelled to some extent in models of ``large'' extra dimensions \cite{large,Arkani,Randall}. Some theories interpreted the brane as a domain wall separating distinct extra-dimensional vacua \cite{wall}, a structure explored in (4+1)D lattice computations \cite{Kaplan,Blum}; others extended to multiple branes and even to a crystalline structure of branes \cite{Kaloper}. A single extra dimension, originally meant to unify gravity with the electromagnetic field has turned insufficient to lodge the growing panoply of particles and fields. This prompted  Pauli \cite{Pauli} to explore two extra dimensions accommodating SU(2) gauge fields an a work that remained unpublished, as he had not arrived at desired results. Lattice methods have been also applied to a (4+2)D model \cite{6d}, where the 4D world is interpreted as a vortex defect in 6D. This topology has been explored in connection with SM phenomenology (\cite{Libanov} and later works by the same group). The number of extra dimensions swells to six \cite{Scherk} or seven \cite{Witten} in string theories. 

The mechanism to be presented here starts tentatively from a 6D manifold and aspires to to understand  the so-called ``Big Bang'' (a misnomer coined by Fred Hoyle as a jibe) as an event of separation between the observable 4D manifold  and remaining two dimensions of a primordial higher-dimensional space. The former is viewed as two coupled spinor manifolds, merged into the Minkowski space when empty and further developing through a symmetry breaking \emph{crystallization} mechanism a structure fitting that of the Standard Model. 

\textbf{Newman -- Penrose formalism.}. Since matter fields are \emph{fermions}, it is natural to describe them as perturbations of spin bases, and to carry out a multiscale perturbation expansion revealing their interactions in the framework of the Newman -- Penrose (NP) formalism. Perturbations, assumed, for the purpose of further analysis, to be weak, matching a suitable ratio of the Planck to phenomenological scales $\epsilon^n \ll 1$ with a suitable exponent $n>0$. 

The starting point is the Minkowski null tetrad expressed in a standard way through base spinors $o^A,\, \iota^A,\,o^{A'},\, \iota^{A'}$ defined on the two spinor manifolds and treated according to the NP formalism. The flat space is spanned by the Minkowski spin bases 
\begin{equation}
\eta_{(0)}^A=o^A, \;\; \eta_{(1)}^A=\iota^A, \;\; \eta_{(0)A}=o_A,\;\; \eta_{(1)A}=-\iota_A. 
 \label{oi}  \end{equation}
and the respective primed (cospinor) bases $\eta{}_{(1)}^{A'}=o^{A'}$ etc. Perturbations of these bases can be either variable or constant on the Planck scale, which will be further associated with the strongly or weakly interacting particles, respectively. 

The basic element of the NP formalism is the array of \emph{spin coefficients} $\gamma_{(ab'cd)}=\eta^A_{(c)}\nabla_{(ab')}\eta_{A(d)}$, where $\nabla_{(ab')}=\eta_{(a)}^A\eta_{(b')}^{B'} \nabla_{AB'}$ are \emph{intrinsic derivatives} commonly denoted as
\begin{equation}
D= \nabla_{(00')}, \;\; \delta = \nabla_{(01')}, \;\; \delta' = \nabla_{(10')},\;\; D'= \nabla_{(11')}.
 \label{DD}  \end{equation}
The common (NP) notation for spin coefficients is (omitting parentheses)  
\begin{align}
&\bs{\kappa}=\gamma_{00'00}, \;\; \bs{\varepsilon} =\gamma_{00'10} , \;\;
\bs{\gamma'}=\gamma_{00'01},\;\;  \bs{\tau'}=\gamma_{00'11} \cr
&\bs{\sigma}=\gamma_{01'00}, \;\; \bs{\beta} =\gamma_{01'10} , \;\;
\bs{\alpha'}=\gamma_{01'01},\;\;  \bs{\rho'}=\gamma_{01'11} \cr
&\bs{\rho}=\gamma_{10'00}, \;\; \bs{\alpha} =\gamma_{10'10} , \;\;
\bs{\beta'}=\gamma_{10'01},\;\;  \bs{\sigma'}=\gamma_{10'11} \cr
&\bs{\tau}=\gamma_{11'00}, \;\; \bs{\gamma} =\gamma_{11'10} , \;\;
\bs{\varepsilon'}=\gamma_{11'01},\;\;  \bs{\kappa'}=\gamma_{11'11}.
 \label{4521}  \end{align}
The priming sign corresponds to the interchange of all 0 and 1 indices. Complex conjugation is equivalent to priming/unpriming of all indices, so that the respective spin coefficients will be further specified as $\gamma_{(ab'c'd')}=\eta^{A'}_{(c')}\nabla_{(ab')}\eta_{A'(d')}$. The deceptively simple Einstein equations are presented in the the NP formalism as an explicit set of 18 equations for spin coefficients and elements of Ricci and Weil tensors $R_{abc'd'}$. For example, the first equation of the NP system reads in the above notation as
\begin{align}
&\nabla_{00'}\gamma_{10'00} - \nabla_{10'}\gamma_{00'00} =|\gamma_{10'00}|^2 
 +\gamma_{01'00}\gamma_{01'00'} \cr
&-\gamma_{00'00}(\gamma_{00'11} +2\gamma_{10'10}
+\gamma_{0'1'1'0'} -\gamma_{10'01}) \cr
& +\gamma_{10'00}(\gamma_{00'10}+\gamma_{0'01'0'})+R_{000'0'},
 \label{l411}  \end{align}
the last term denoting a component of the Ricci tensor.

\textbf{Waves in spin-spaces}.The flat vacuum state \eqref{oi} will be further perturbed by weak gravitational waves with momenta $\mathbf{k}_a=\kappa_A\kappa_{A'}$ measured on the Planck scale: 
\begin{align}
 &\qquad \eta_{(a)}^A \to \eta_{(a)}^A  + \epsilon\sum_f\zeta^A_{(af)} \mathcal{E}_{(f)}, 
  \label{zeta}  \\
&\mathcal{E}_{(f)}= \exp\left(\I K_{j(f)}K'_{j'(f)}X^jX^{j'}\right) 
 \equiv \exp\left(\I k_{j(f)}x^j\right).
   \label{exp}  \end{align}
Here $x^j,  k_{j(f)}$ are Planck-scale coordinates and momenta, respectively, expressed  through appropriate spinors $X^j,  K_{j(f)}$ and cospinors $X^{j},  K'_{j'(f)}$. The primed (cospinor) bases are perturbed in a similar way, with the primed indices $A',a'$.

Perturbations of the derivative operators \eqref{DD} originate in perturbations of associated spin bases:
\begin{eqnarray}
\nabla^{[1]}_{(ab')}  &=  & \epsilon \sum_{f,g}\left(\eta_{(a)}^A \zeta_{(b'f)}\mathcal{E}_{(f)}
 +\eta_{(b')}^{B'} \zeta_{(ag)}^A \mathcal{E}_{(g)}\right)  \nabla_{AB'},\cr
 \nabla^{[2]}_{(ab')}  &= &  \epsilon^2\sum_{f,g} \zeta_{(af)}^A  \zeta_{(b'g)}^{B'}
  \mathcal{E}_{(f)}\mathcal{E}_{(g)} \nabla_{AB'}.
 \label{DD1}  \end{eqnarray}
As the exponent in Eq.~\eqref{exp} contains the sum of all coordinates and all spin frames may oscillate with the same set of momenta, the action of $\nabla_{(ab')}$ is the same for all $a,b'$ and just reduces to multiplication by $-\mathbf{k}_f$, so that the indices can be omitted. 

Since $ \nabla_{(ab')}$ act nontrivially only on perturbed bases, all spin coefficients are at most of  $\mathcal{O}(\epsilon)$. In the first order, 
\begin{equation}
 \gamma^{[1]}_{(cd)}=  -\epsilon \eta^{A}_{(c)}
 \sum_{f} \mathbf{k}_f \zeta_{A(df)} \mathcal{E}_{(f)}.
 \label{gam1}  \end{equation}
In the second order, there are two distinct expressions:
\begin{eqnarray}
&&\gamma^{[2]}_{(cd)}= \nabla \gamma^{[1]}_{(cd)}=\epsilon^2 \sum_{f,g}
\zeta^{A}_{(cf)}\zeta_{A(dg)}\mathbf{k}_f \cdot \mathbf{k}_g 
 \mathcal{E}_{(g)} \mathcal{E}_{(f)},   \label{gam2a}\\
  &&  \gamma^{[2]}_{(ab'cd)}=\eta_{(c)}^A\nabla^{[1]}_{ab'}\eta_{A(d)}=  \cr
 && \epsilon^2 \eta^A_{(c)} \nabla_{AB'}\sum_{f} \left(\eta_{(a)}^C \zeta_{(b'f)}\mathcal{E}_{(f)}
 +\eta_{(b')}^{B'} \zeta_{(af)}^C \mathcal{E}_{(f)}\right).
  \label{gam2b}  \end{eqnarray}
  
The linear $\mathcal{O}(\epsilon)$ terms in the left-hand side of  Eq.~\eqref{l411} contain intrinsic derivatives acting upon the perturbations of  bases denoted by the last number in the definition of the respective spin coefficient. Taking Eq.~\eqref{l411}, we see that the last two numbers in the indices of the two spin coefficients are identical, while the indices of the intrinsic derivative acting on one spin coefficients is identical to the first two indices of another one, so that their combination vanishes.The same applies to pairs of linear terms in the rest of NP equation system. 

A wave with a Planck-scale momentum is a \emph{ghost} that averages to zero on phenomenological scales. Only constant terms, which may be subject to long-scale modulation, are relevant for observable phenomena. Such terms may arise in $\mathcal{O}(\epsilon^2)$ only as \emph{standing waves}. The linear terms in the $\mathcal{O}(\epsilon^2)$ NP equations contain unperturbed directional derivatives $\nabla$ acting upon $ \gamma^{[2]}_{(cd)}$ and $\nabla^{[1]}_{ab'}$  directional derivatives acting upon $\gamma^{[1]}_{(cd)}$. The former are equivalent to $\gamma^{[2]}_{(cd)}$ in  Eq.~\eqref{gam2a} and generate a standing wave when $\mathbf{k}_f = - \mathbf{k}_g$. Returning to Eq.~\eqref{l411}, we see that since this expression is independent of $(ab')$ both linear terms cancels, as they do in $\mathcal{O}(\epsilon)$. However, standing waves can be generated in the second order of the multiscale expansion by nonlinear terms combining terms of $\gamma^{[1]}_{(cd)}$ with momenta of opposite signs. 

\textbf{Strong interactions}. The perturbations amplitudes of spinor and cospinor bases, are interpreted as \emph{quarks}, with the index $a$ denoting their \emph{flavor} and the index $f$ related to their \emph{color}, as defined below. By the construction from a spin space, the quarks are \emph{fermions}. The quarks originating from the same subspace, (spinor, cospinor, or passive) are viewed as belonging to the same \emph{generation}.  

The two extra dimensions additional to the Minkowski world space may be symmetric when unperturbed, and, possibly, spanned by the two hidden time variables. 
It is reasonable to assume these ``extra-world'' perturbations to be \emph{passive}, i.e., driven exclusively by 4D perturbations outlined above, and defined exactly as in Eq.~\eqref{zeta} with $\eta_a, \,\zeta_{(af)}$ being spinors defined on the passive 2D manifold. Accordingly, only the world derivative operators \eqref{DD} and their perturbations \eqref{DD1} are relevant. An analog of such a passive excitation in (2+1)D, far easier to visualize than (4+2)D, is excitation of waves in a fluid in a vicinity of an oscillating 2D membrane. The absence of independent activity in a fluid allows in this model case to derive it from the membrane dynamics. By analogy, we can concentrate on the 4D gravity in the description in the (4+2)D world.  as well as bases of the passive 2D manifold $\zeta^A_{(af)}, \, \zeta^{A'}_{(a'f)}$ in Eq.~\eqref{zeta}

This primordial ``meson world'' supported by second-order interactions would arise and set off expanding, thereby justifying the local transition from a formless 6D manifold, provided its energy, (which could be computed by solving the full set of NP equations) is below the zero vacuum level. 

As an incipient universe expands, third-order interactions including quark triplets come into the play. Planck-scale oscillations are suppressed in triplet combinations with momenta $\mathbf{k}_f$ forming a resonant triangle. Relevant solutions, containing only a restricted set of wavenumbers $k_{(f)}$, should include \emph{resonant} triplets of waves in the spinor, cospinor, and passive 2D manifolds adding up to zero, in the same way, as common macroscopic 2D and 3D patterns combine resonant combinations of waves. This is possible only if all momenta $k_{(f)}$ are spacelike in a certin spacetime frame. A resonant triangle does not need to be equilateral, which contrasts a common situation in macroscopic symmetry breaking phenomena \cite{ch93,lp06,alx}. 

A vacuum crystal would naturally arise which is the overall energy is further reduced due to triplet resonances in the third-order of the expansion of the NP equations. The three modes of the resonant triangle are interpreted as \emph{colors}. The ``white'' state combining all three colors is necessary for suppressing third-order oscillations. Assigning a particular color to constituent modes of a resonant triangle is arbitrary, and it is not necessary to attribute color exchange and ``color confinement'' to interactions meditated by gluons. 

Each quark comes in three colors; antiquarks relate to reverse waves.
The ``color symmetry'' ensures satisfying the resonance condition in triplet interactions involving any set of quarks, and therefore quark triplets combined in \emph{baryons} are non-separable. \emph{Mesons} combining a quark and an antiquark (and therefore colorless) are recognized as standing spinor waves in the second order of the above expansion underlying the emergence of the vacuum crystal. In this framework, there is no need  to attribute quark confinement to a vague and counterintuitive notion of interaction strength increasing with separation. 

As it has been briefly discussed in the introduction, higher-dimensional theories have a long history but have never found either experimental evidence or even an unequivocal theoretical justification. Six-dimensional spinors have been invoked both as a device for unification of Standard Model fields and as the background of supersymmetric theories \cite{6d}. In Wheeler's vision,``the dynamics of Einstein's curved space geometry runs its course in superspace as the dynamics of a particle unfolds in spacetime'' \cite{wh64}. The above scheme envisages the ``Big Bang'' as a combination of separation between the 4D manifold oriented as the common Minkowski space and the passive extra-time manifold, wuth the former further developing into an observable universe.

\textbf{Discussion}. The key point of the current proposal is an ultimate unification presenting matter fields as nothing more than modulated Planck-scale gravitational waves. The symmetry should be restored within black holes, as crystalline units dissolve when space contracts beyond the limit allowing for a multiscale analysis. Although the theory is purely ``classical'', it does not call for conventional quantization but rather for a reinterpretation of quantum mechanics attributing the source of quantum uncertainty to the envelope character of phenomenological fields, in line with various proposals postulating a foamlike structure of spacetime \cite{g98} (first suggested
by Wheeler \cite{wh64}), the existence of a minimal length \cite{Elze}, an atomistic spacetime,  \cite{Kempf,min}, or a Planck-scale periodicity \cite{Jizba, Goshir} (unrelated to SM).

It is pointless to speculate on the existence of a ``metaphysics'' below the Planck scale, that would determine resonant momenta. The proposed theory does not predict parameters of SM dependent on the parameters of the underlying crystal, which should have been selected at ``Big Bang''. It is likely that different resonances are responsible for crystalline structures in different universes that may arise at unconnected locations within the underlying 6D manifold, and the resonant momenta may be inferred from observed quark masses rather than the other way around. 

In principle, it might be possible to compute the energy gain due to crystallization \emph{ab initio} by solving the NP equation system to the third order and computing the Ricci tensor that determines the energy of crystals based on a specific resonances. Still, there is no warranty that the ``optimal'' crystalline structure would be chosen. The anthropic principle may be evoked to estimate values appropriate to a long-lived universe and fitting the observed phenomenology. Still less feasible would be computing quark masses \emph{ab initio} by incorporating Eq.~\eqref{gam1} in the multiscale expansion, but distinctions between flavors can be naturally attributed to an asymmetry of the vacuum lattice.

The inferred ``inflation'' period corresponds to a rapid growth of a fluctuation in a primordial flat space developing into a sustainable crystal. A small assembly of cells, likely disordered, is not a crystal yet. A three-wave resonance leading to the genesis of quarks emerges at the size of $\mathcal{O}(\epsilon^{-3})$ Planck units. Identifying this with the ``grand unification'' epoch dominated by quark--gluon plasma at $10^6$ seconds ($t \sim 10^{39}$ Planck units) after the Big Bang suggests the scaling estimate $\epsilon \sim 10^{-13}$. 

The crystalline scale remains invariant as a universe expands: more crystalline units are just added to fill the available space within the 6D manifold. The energy gain due to crystallization can be identified with the ``dark energy'' driving the expansion of a universe. The expansion would slightly accelerate, as it actually does, when the boundary of  a universe in the 6D manifold extends, similar to surface tension effects leading to Ostwald ripening \cite{ost} in macroscopic two-phase systems. ``Dark matter'' may be built up by macroscopic fields that are unrelated to the Planck-scale crystal and therefore do not interact with its modulations. In an unlikely case of a collision between two universes based on different resonant momenta, the two modulating fields may not interact remaining mutually dark, unless, by chance, a new resonance is formed, leading to a more complicated, and likely disastrous, phenomenology.



\begin{thebibliography}{9}
\bibitem{ch93}M.C. Cross and P.C. Hohenberg, Pattern formation outside of equilibrium, Rev. Mod. Phys. \textbf{65}, 851 (1993).
\bibitem{lp06}L.M. Pismen, Patterns and Interfaces in Dissipative Dynamics, Springer, Berlin, 2006;
2nd edition Springer, Cham, 2023.
\bibitem{kk1}M.J. Duff, B.E.W. Nilsson, and C.N. Pope, Kaluza-Klein supergravity, Phys. Rep.  \textbf{130} 1 (1986).
\bibitem{large}I. Antoniadis, A possible new dimension at a few TeV, Phys. Lett. B \textbf{246}, 377 (1990).
\bibitem{Arkani}N. Arkani-Hamed, S. Dimopoulos S, and G. Dvali, The hierarchy problem and new dimensions at a millimeter, Phys. Lett. B \textbf{429}, 263 (1998)
\bibitem{Randall}L. Randall and R. Sundrum, Large Mass Hierarchy from a Small Extra Dimension, Phys. Rev. Lett. \textbf{83}, 3370 (1999).
\bibitem{wall}V.A. Rubakov and M.E. Shaposhnikov, Do we live inside a domain wall? Phys. Lett. B \textbf{125}, 136 (1983).
\bibitem{Kaplan}D.B. Kaplan, A method for simulating chiral fermions on the lattice, Phys. Lett. B \textbf{288}, 342 (1992).
\bibitem{Blum}T. Blum and A. Soni, QCD with domain wall quarks, Phys. Rev. D \textbf{56}, 174 (1997).
\bibitem{Kaloper}N. Kaloper, Crystal manyfold universes in AdS space, Phys. Lett. B, \textbf{474}, 269 (2000). 
\bibitem{Pauli}L. O'Raifeartaigh and N. Straumann, Gauge theory: Historical origins and some modern developments, Rev. Mod. Phys.  \textbf{72}, 1 (2000).
\bibitem{6d}F. Bauer, T. Haellgren, and G. Seidl, Discretized gravity in 6D warped space, Nucl. Phys. B \textbf{781}, 32 (2007).
\bibitem{Libanov}M.V. Libanov and S.V. Troitsky, Three fermionic generations on a topological defect in extra dimensions, Nucl. Phys. B \textbf{599} 319 (2001).
\bibitem{Scherk}J. Scherk and J.H. Schwarz, Dual field theory of quarks and gluons, Phys. Lett. B \textbf{57} 46 (1975).
\bibitem{Witten}E. Witten, Search for a realistic Kaluza-Klein theory, Nucl. Phys. B \textbf{186}, 412 (1981).
\bibitem{NP}E. Newman and R. Penrose, An Approach to Gravitational Radiation by a Method of Spin Coefficients, J. Math. Phys. \textbf{3}, 566 (1962).
\bibitem{PR}R. Penrose and W. Rindler, Spinors and Space-Time, Cambridge University Press, 1987.
\bibitem{alx}S. Alexander and J. McTague, Should All Crystals Be bcc? Landau Theory of Solidification and Crystal Nucleation, Phys. Rev. Lett. \textbf{41}, 702 (1978).
\bibitem{6d} D. Chester, A. Marrani, and M. Rios, Beyond the Standard Model with Six-Dimensional Spinors, Particles, \textbf{6}, 14 (2023).
\bibitem{wh64} J.A. Wheeler, Geometrodynamics and the issue of the final state, Relativity, Groups and Topology, p. 509, Gordon and Breach (1964); C.W. Misner, K.S. Thorn, and J.A. Wheeler, Gravitation, W.H. Freeman and Co. (1973).
\bibitem{g98}L.J. Garay, Spacetime Foam as a Quantum Thermal Bath, Phys. Rev. Lett. \textbf{80}, 2508 (1998).
\bibitem{Elze}H.-T. Elze, Does quantum mechanics tell an atomistic spacetime? J. Phys. Conf. Ser. \textbf{174}, 012009 (2009)
\bibitem{Kempf}A. Kempf, G. Mangano, and RB. Mann, Hilbert space representation of the initial length uncertainty relation, Phys. Rev. D \textbf{52}, 1108 (1995).
\bibitem{min}S. Hossenfelder, Minimal length scale scenarios for quantum gravity, Living Rev. Relativity, \textbf{16}, 2 (2013).
\bibitem{Jizba}P. Jizba, H. Kleinert, and F. Scardigli, Uncertainty relation on a world crystal and its applications to micro black holes, Phys. Rev. D \textbf{81}, 084030 (2010).
\bibitem{Goshir}S. Goshir, Explanation of the generalizations of uncertainty principle from coordinate and momentum space periodicity, Eur. Phys. J. Plus 139-569 (2024). 
\bibitem{ost}W. Ostwald, Studien \"uber die Bildung und Umwandlung fester K\"rper, Z. Phys. Chem. \textbf{22}, 289 (1897).
\end{thebibliography}
\end{document}